\documentclass[aps,prl,twocolumn,showpacs,a4paper]{revtex4}

\usepackage{graphicx}

\begin{document}

\setlength\arraycolsep{1pt}

\title{Domain-Wall Waves (2D Magnons) in Superconducting Ferromagnets}

\author{N.~A.~Logoboy}
\author{E.~B.~Sonin}
\affiliation{Racah Institute of Physics, The Hebrew
University of Jerusalem, Jerusalem 91904, Israel}
\date{\today}
\begin{abstract}
Propagation of the magnetization waves along domain walls (2D
magnons) in a superconducting ferromagnet has been studied
theoretically. The magnetostatic fields (long-range dipole-dipole
interaction) have a crucial effect on the spectrum of 2D magnons. But
this effect is essentially affected by the superconducting Meissner
currents, which screen the magnetostatic fields and modify the
long-wavelength spectrum from square-root to linear. The excitation
of the domain wall waves by an electromagnetic wave incident on a
superconducting-ferromagnet sample has been considered. This suggests
using measurements of the surface impedance for studying the domain
wall waves, and eventually for effective probing of
superconductivity-ferromagnetism coexistence.
\end{abstract}


\maketitle

Coexistence of superconductivity and ferromagnetism, which results
in a number of unusual phenomena, has already been studied about 50
years \cite{ginz}. A revival of research in this area was stimulated
by experimental observation of superconductivity-ferromagnetism
coexistence in high-$T_c$ superconductors \cite{Felner,bernhard} and
various unconventional superconductors
\cite{Eskildsen,saxena,pfleiderer,aoki}. An essential obstacle for
experimental detection of superconductivity-ferromagnetism
coexistence is screening of internal magnetic fields generated by
the ferromagnetic order parameter (spontaneous magnetization) by
superconducting currents. A possible way to overcome this obstacle
is to investigate spin waves, which are a direct evidence of the
presence of spontaneous magnetization \cite{NV,Braude}. They can be
excited by the electromagnetic (EM) wave incident on a
superconducting ferromagnet (SCFM).

In Ref. \onlinecite{Braude} spin waves in SCFMs have been studied
for the Meissner state with the uniform spontaneous magnetization,
i.e., for a single-domain sample. Though  at the equilibrium
superconductivity suppresses usual ferromagnetic domain structures
with periods determined by demagnetization factors and sizes of
samples \cite{Sonin}, the interplay with superconductivity can lead
to formation of intrinsic domains of the order of the London
penetration depth or less \cite{krey} (see the latest discussion in
Ref.~\onlinecite{Buzdin}). Moreover, a domain wall (DW) is a
topologically stable defect, and the presence (absence) of DWs can
be determined by the sample prehistory but not by the conditions of
the equilibrium. This justifies an interest to studying spin waves
propagating along DWs (2D magnons).

In normal ferromagnet (FM) the DW dynamics and DW waves have already
been studied a few decades and are important for various
applications \cite{Malozemoff}. The crucial feature of 2D magnons is
that  the magnetostatic effects (dipole-dipole interaction) are much
more important for them than for 3D bulk magnons. On the other hand
one may expect that SC should influence these effects first of all
since it prohibits penetration of the magnetostatic fields deep into
the bulk of domains.

In the present Letter we investigate theoretically DW waves in a
SCFM. The analysis confirms expectation of strong effect of
superconductivity on DW waves: Meissner screening of magnetostatic
fields in domains modifies the square-root 2D-magnon spectrum
revealed for normal FM \cite{Malozemoff} to a linear sound-like
spectrum.  We show how the 2D magnons can exhibit themselves in the
linear response to the AC magnetic field (surface impedance). As
well as in the case of bulk 3D spin waves, detection of 2D magnons,
being interesting itself, would provide a direct probe of
superconductivity-ferromagnetism coexistence. The spectrum of 2D
magnons is sensitive to bulk and surface pining of DWs, which can
lead to a gap in the spectrum and a peak in the EM wave absorption.

\begin{figure}[t]
  \includegraphics[width=0.4\textwidth]{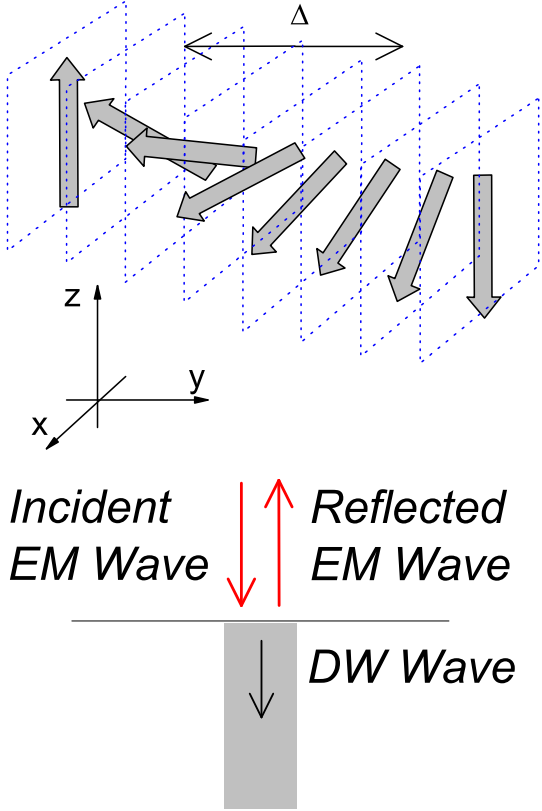}
\caption{(Color online)
(a) The $180^\circ$
Bloch domain wall. The magnetizations in the
domains are parallel and antiparallel to the $z$ axis. Inside the DW
the magnetization rotates in the $xz$ plane. (b) Excitation of DW
waves by an incident EM wave.}
\label{Setup}
\end{figure}

Let us consider a $180^\circ$ DW in a FM with magnetic anisotropy of
the easy-axis type [Fig.~\ref{Setup}(a)]. The $z$ axis is the easy axis
and the DW is parallel to the $xz$ plane and separates two domains
with spontaneous magnetization $\mathbf{M}$ along the $+z$ ($y<0$) and
$-z$  ($y>0$) directions. Neglecting the magnetostatic fields, the
structure of the DW is well known \cite{LL,Malozemoff}. Rotation of
the magnetization $\mathbf{M} = M_{0}(\cos\phi\ \sin\theta,\sin\phi\
\sin\theta,\cos\theta)$ inside the DW is described in polar and
azimuthal angles $\theta$ and $\phi$ as \cite{LL}
\begin{eqnarray}
\theta_{0}=2\tan^{-1} e^{y/\Delta}\, .
 \label{eq:ground state}
\end{eqnarray}
Here $\Delta =(A/K)^{1/2}M_{0}$ is the DW width, $A$ is the
constant, which determines the energy of the non-uniform exchange $A
\nabla_j \mathbf{M} \cdot  \nabla_j \mathbf{M}$, and the constant
$K$ determines the easy-axis anisotropy energy $K \sin^{2} \theta$.
The azimuthal angle $\phi$ is an arbitrary constant as far as the
magnetostatic (dipole) fields are neglected. In normal FMs currents
are absent at the equilibrium , and the magnetic field
$\mathbf{H}=\mathbf{B}-4\pi\mathbf{M}$ is curl-free, i.e., is a
potential field with sources  called ``magnetic charges'' $\rho_m=-
\mathbf{\nabla}\cdot \mathbf{M}$:
$\mathbf{\nabla}\cdot\mathbf{H}=\mathbf{\nabla}\cdot\mathbf{B}-4\pi
\mathbf{\nabla}\cdot\mathbf{M}=4\pi \rho_m  $. The energy of the
magnetostatic fields lifts degeneracy with respect to the angle
$\phi$: the ground state corresponds to the DW of the Bloch type
with $\phi=0$. This means that the magnetization rotates in the DW
plane, and the magnetic charges do not appear ($\mathbf{\nabla}\cdot
\mathbf{M}=0$). This structure, which was obtained for normal FMs,
remains valid at scales of order $\Delta$ also for SCFMs, as far as
the London penetration depth $\lambda$ essentially exceeds the DW
thickness $\Delta$. However, the difference between a normal FM and
a SCFM is important at distances larger than $\Delta$: while in
normal FMs the magnetic field $\mathbf{H}=\mathbf{B}-4\pi\mathbf{M}$
vanishes and the magnetic induction $\mathbf{B}= 4\pi\mathbf{M}$ is
constant inside domains, in SCFMs the magnetic induction
$\mathbf{B}$ is confined in the Meisssner layers of width $\lambda$
\cite{Sonin}:
\begin{eqnarray}
B_z= \pm 4\pi M_0 e^{\pm y/\lambda}~,
 \end{eqnarray}
where the upper and the lower signs correspond to $y<0$ and
$y>0$ respectively. Thus the Meissner currents
$j_{x}=-(c/4\pi) dB_z/dy$ screen out the main  bulk of domains
from the magnetic induction.

We shall use the DW dynamics developed for FMs with high quality
factor $\alpha =K/2\pi M^{2}_{0}$ \cite{Malozemoff}. In this limit
one may assume that the structure of the Landau-Lifshitz DW, which
is given by Eq.~(\ref{eq:ground state}), is not affected essentially
by dynamical processes, and the DW dynamics can be described in the
terms of the pair of canonically conjugated variables: the azimuthal
angle $\phi(x,z;t)$, which determines the plane of magnetization
rotation, and the displacement of the DW $\eta=\eta(x,z;t)$ along
the $y$ axis: $\theta(x,y,z;t)=\theta_{0}(y-\eta(x,z;t))$.
Neglecting dissipation, the equations of DW motion (Slonczewski's
equations \cite{Slonczewski}) are the Hamilton equations:
\begin{eqnarray}
\label{eq:Slonczewski}
\partial_{t}\eta=\frac{\gamma}{2M_{0}}\frac{\delta H}{\delta\phi}~,~~~
\partial_{t}\phi=-\frac{\gamma}{2M_{0}}\frac{\delta H}{\delta\eta},
\end{eqnarray}
where $\delta/\delta \phi$ and $\delta/\delta \eta$ are functional
derivatives and the Hamiltonian $H=\int \sigma dx\,dz$ is determined
by the surface energy density $\sigma$ of the DW. The Hamilton
equations (\ref{eq:Slonczewski}) point out a direct analogy of
$(2M_0/\gamma) \phi$ with a canonical momentum conjugate to
``coordinate'' $\eta$. But it is worthwhile to mention another
analogy. The displacement $\eta$ leads to the change $2M_0\eta$ of
the $z$-component of the total magnetization per unit area  of the
DW. Since $\phi$ is the angle of spin rotation around the $z$ axis,
Eqs.~(\ref{eq:Slonczewski}) can be also considered as Hamilton
equations for the pair of conjugate variables ``angular
momentum--angle''.

The DW surface energy is
\begin{eqnarray}
\label{eq:surface energy density}
\sigma(\phi,\eta)= 4\pi M^{2}_{0}\Delta \alpha  \left[
(\mathbf{\nabla}\eta)^{2}+\Delta^{2}
(\mathbf{\nabla}\phi )^{2}\right]
\nonumber \\
+\frac{1}{8\pi}\int dy (\mathbf {h}^2 +\lambda^{2}[\mathbf{\nabla}
\times \mathbf {h}]^{2})
 \nonumber\\
+\frac{1}{4\pi}\int dy (\mathbf H^{(0)}\cdot \mathbf
{h }_2+\lambda^{2}[\mathbf {\nabla}
\times \mathbf H^{(0)}]\cdot [\mathbf {\nabla}
\times \mathbf {h}_2])~. \qquad
              \end{eqnarray}
The terms $\propto \alpha$ (the first square brackets) present  the
contributions of the exchange energy and the anisotropy energy,
whereas the integral terms are the energies of superconducting
screening currents and of the magnetostatic fields $\mathbf
{H}=\mathbf H^{(0)}+ \mathbf {h}+ \mathbf {h}_2$. Here $\mathbf H^{(0)}$
is the magnetic field in the ground state, and $\mathbf {h}$ and
$\mathbf {h}_2$ are the dynamic corrections to the field of the
first and the second order in the wave amplitude respectively.
Necessity to take into account the second-order corrections is a
peculiar feature of the DW dynamics in SCFM: in normal FM
$\mathbf H^{(0)}=0$ and these corrections vanish.

The first-order contribution to the magnetic
induction $\mathbf b=\mathbf B-\mathbf B^{(0)}=\mathbf h+4\pi \mathbf
m$ can be found from the generalized London equation:
\begin{equation}
\label{eq:London} -\Delta ~ \mathbf b+ \lambda^{-2}\mathbf
b=4\pi \mathbf {\nabla} \times \mathbf
{\nabla} \times \mathbf m. \\
\end{equation}
The dynamic components of the magnetization $\mathbf {m}=\mathbf
M-\mathbf M^{(0)}$ are defined through the conjugated variables $\eta$
and $\phi$  as follows:
\begin{equation} \label{eq:m}
\mathbf{m}=M_{0} \left( \begin{array}{ccc}
-\eta\,d [\sin{\theta_{0}(y)}]/dy  \\
\phi\sin{\theta_{0}(y)} \\
-(\eta/\Delta)\sin^{2}{\theta_{0}(y)}
\end{array} \right).\\
\end{equation}

Let us consider the plane wave with wave vector $\mathbf k =(0,0,k)$
parallel to the $z$ axis: $\phi, \eta, \mathbf{b} \propto \exp(-i\omega
t+ikz)$. Then the London equation (\ref{eq:London})
becomes a set of ordinary differential equations with the only
coordinate $y$, and its solution  is:
\begin{eqnarray}
\mathbf b = - \frac{4
\pi}{\widetilde{k}}\left[
e^{\widetilde{k}y}\int_{y}^{\infty} dy'\mathbf {\nabla}
\times \mathbf {\nabla} \times \mathbf m(y')e^{-\widetilde{k}y'}  \right.
\nonumber \\ \left.
 +e^{-\widetilde{k}y}\int_{-\infty}^{y} dy'\mathbf {\nabla} \times \mathbf
{\nabla} \times \mathbf m(y')e^{\widetilde{k}y'}\right].
 \label{eq:magnetic induction}
\end{eqnarray}
Here $\widetilde{k}^{2}=k^{2}+\lambda^{-2}$.

Now one can substitute the magnetic field
$\mathbf{h}=\mathbf{b}-4\pi\mathbf{m}$ into the first magnetostatic
integral in Eq.~(\ref{eq:surface energy density}) and perform
integration. The contribution of the second-order corrections
$\mathbf {h}_2$ [the second magnetostatic integral in
Eq.~(\ref{eq:surface energy density})]can be determined from general
arguments. The second-order terms contain the Fourier components of
$\eta^2$ with zero wave vector and with the wave vector $2k$ (the
second harmonic). Only the first component, which is
$z$-independent, can interfere with the zero-order terms. This
yields the $k$-independent contribution to the surface energy, which
should provide translational invariance: the surface energy should
not depend on the spatially independent DW displacement. Altogether
this yields the following surface energy density in variables $\eta$
and $\phi$:
\begin{eqnarray}
\label{eq:effective surface energy density}
   \sigma=4\pi M^{2}_{0}
   \left [\alpha k^2 \Delta (\eta^2 +\phi^2 \Delta^2) \right. \nonumber\\
    +\left.\left(1-\frac{\Delta k^2}{ \arrowvert\widetilde {k}\arrowvert}\right)
\phi ^{2}\Delta+\left(\arrowvert \widetilde {k}\arrowvert-{1\over
\lambda}\right)\eta^{2}
 \right ]~,
\end{eqnarray}
where the second-order terms are presented by the negative term
$-\eta^2/\lambda$. All terms, which are not proportional to the
quality factor $\alpha$, originate from the magnetostatic energy and
the SC currents. The magnetostatic term $\propto \phi^2$ plays a
role of the ``kinetic'' energy since the angle $\phi$ is the
momentum conjugate to the displacement $\eta$. In the limit $k \to
0$ it is determined by the magnetic fields inside the DW and can be
expressed via the D\"oring mass $m_D =1/\gamma^{2}2\pi \Delta$
\cite{Malozemoff}. The $k$-dependent contribution to the
kinetic energy represents the energy of the magnetostatic fields
outside the DW in the layer of width $\sim 1/|\tilde k|$. Altogether
the magnetostatic terms $\propto \phi^2$ are connected with the
double layer of magnetic charges, which appears when the DW rotates
from the Bloch orientation $\phi =0$. On the other hand, the
magnetostatic term $\propto \eta^2$ is connected with the magnetic
charges, which appear due to rotation of the DW with respect to the
direction of the magnetization in domains.

Using Eq.~(\ref{eq:effective surface energy density}) in Slonczewski's
equations (\ref{eq:Slonczewski}) transformed to the Fourier presentation
one obtains the spectrum of plane waves propagating along
the DW:
\begin{eqnarray} \label{eq:long wave length}
\omega^{2}=\omega_{M}^2\left[\alpha k^{2}\Delta^{2}+{\Delta\over
\lambda}\left(\sqrt
{k^{2}\lambda^{2}+1}-1\right)  \right] \nonumber \\
\times \left(1+\alpha k^{2}\Delta^{2}-\frac
{k^{2}\lambda\Delta}{\sqrt {k^{2}\lambda^{2}+1}}\right),
\end{eqnarray}
where $\omega_{M}= 4\pi \gamma M$.

In the limit  $\lambda \to \infty$ (no superconductivity), this
transforms to the spectrum of 2D magnons in normal FMs:
\begin{equation} \label{eq:FM}
\omega^2=\omega_M^2(\alpha k^{2}\Delta^{2}+\arrowvert k
\arrowvert\Delta)(1+\alpha k^{2}\Delta^{2}- \arrowvert k
\arrowvert\Delta).
\end{equation}
All nonanalytic terms $\propto |k|$ in this expression are of
magnetostatic origin and are related with penetration of
magnetostatic fields deep into domains on scales of the order of the
wavelength $2\pi/k$. They lead to the nonanalytic wave spectrum
$\omega \propto \sqrt{k}$ in the long-wavelength limit, which was
known before \cite{Malozemoff}. Another nonanalytic term, which
appears in the second multiplier with the negative sign,  has not
been considered so far as far as we are aware.

Returning back to SCFM one can see that the Meissner screening
eliminates non-analytic features of the spectrum and allows expansion in
$k^2$. Thus in the long-wavelength limit the spectrum is sound-like:
\begin{equation} \label{eq:SCFM}
\omega =c_s k ,
\end{equation}
where $c_s =\omega_M  \sqrt{\Delta(\alpha  \Delta +\lambda/2)}$ is
the 2D spin-wave velocity.

In order to find the effect of DW waves on surface impedance one
should solve a boundary problem. In general the wave spectrum
Eq.~(\ref{eq:long wave length}) leads to
differential equations of high order in the configurational space
(bearing in mind the correspondence $k \to -i
\partial/\partial z$). But the sound-like spectrum Eq.~(\ref{eq:SCFM}) reduces
Slonczewski's equations  in the configurational space  to two ordinary differential
equations of the first order:
\begin{eqnarray}
\label{dif}
\partial_{t}\eta=\omega_M\Delta \phi-\gamma \Delta h_{ext}~,
~~~\partial_{t}\phi={c_s^2 \over \omega_M \Delta}
{\partial^2
\eta\over
\partial z^2}~.\quad
\end{eqnarray}
In  the first equation we added the interaction with the external
magnetic field. This field arises from the  EM wave of frequency
$\omega$ incident on the sample surface and linearly polarized along
$y$-axis [Fig.~\ref{Setup}(b)]. The EM wave penetrates through the
surface of SCFM at the distance $\sim \lambda$:
$h_{ext}=h_0e^{z/\lambda}$. Here $h_0$ is the AC magnetic field at
the sample surface $z=0$.

Next is to formulate boundary conditions for these equations. One
boundary condition is imposed at $z\to -\infty$. We assume that the
DW plane wave, which is excited near the sample surface $z=0$ is the
only propagating wave in the bulk and there is no reflected wave
coming from $z=-\infty$. The second boundary condition is imposed at
the sample border $z=0$. Assuming an ideal surface without any
surface force the balance of the ``momentum'' $\phi$ requires  that
the DW remains normal to the sample border: $\partial \eta/\partial
z=0$. But the magnetic field  $h_{ext}$ generated by the incident EM
wave , though being a bulk force, is confined to the Meissner layer
of width $\lambda$, which is much smaller than the wavelength.
 Then one may consider its integral effect as that of a surface
force. This leads to modification of the boundary condition for
propagating wave, which must be written as
\begin{eqnarray} \label{BC}
 c_s^2 {\partial\eta\over\partial z}\bigg\arrowvert_{z=0} -\frac{\pi}{2} i\omega
\gamma h_0\Delta\lambda=0~.
                                                  \end{eqnarray}
This boundary condition determines the amplitude of the wave propagating
from the sample border to $z= -\infty$:
\begin{equation}
\eta=\frac{\pi}{2}{\omega\over c_s^2 k} \gamma
h_0\Delta\lambda\exp{(-i \omega t-i kz)}
 \label{eq:displacement}~.
\end{equation}
The wave is accompanied by the energy flux, which is determined from the
energy balance $\partial_t\sigma+\partial S/\partial z=0$:
\begin{eqnarray} \label{eq:energy flux}
  S =-{\partial \sigma \over  \partial (\partial \eta/\partial z)}
  \partial_{t}
\eta =-{\pi \over 16  } {\omega^2 \over c_s} \Delta \lambda^2h_0^2~.
        \end{eqnarray}
The energy brought away by the DW plane wave is supplied by the wave
incident on the surface: at the reflection of the wave from the
sample surface some part of energy is absorbed though no dissipation
mechanism is present explicitly in our model: it is assumed that
dissipation occurs deep inside the sample where the spin wave
eventually dissipates and does not return as a reflected wave.
Energy absorption  at reflection of the EM wave from the sample
surface is characterized by real part of the surface impedance,
which is defined as a ratio of the tangential components of electric
field $e_{x}$ and magnetic field $h_{0}$ at the surface: $\zeta=
e_{x}/h_0\arrowvert_{z=0}$. Namely, absorption per unit area per
second is given by $(ch_0^2/8 \pi) \mbox{Re} \zeta$. Equating this
to the average energy flux per unit area given by $\vert S \vert n_{
W}$ one obtains that
\begin{eqnarray} \label{eq:SI final}
 \mbox{Re}\zeta=8 \pi n_{W}{\vert S \vert \over c h_e^2}=\frac{\pi^{2}}{2} n_{W} \Delta \lambda^2{\omega^2 \over c c_s}~.
\end{eqnarray}
Here $n_{W}$ is the linear density of DWs. However this dependence
is essentially modified by bulk and surface pinning of DW. Bulk
pinning lifts translational invariance and leads to the gap in the
magnon spectrum, which instead of Eq.~(\ref{eq:SCFM}) becomes
\begin{equation}
\omega^2 =\omega_{p}^2+c_s^2 k^2~.
\end{equation}
Surface pinning modifies the boundary condition Eq.~(\ref{BC}), which now is
\begin{eqnarray}
 c_s^2 \left({\partial\eta\over\partial z}+k_s \eta\right)\bigg\arrowvert_{z=0}
- i\omega \gamma h_0\Delta\lambda=0~,
                                                  \end{eqnarray}
where $k_s$ is the inverse length characterizing intensity of surface pinning.
Repeating the derivation of $\mbox{Re}\zeta$ in the presence of pinning one obtains
that
\begin{eqnarray}
 \mbox{Re}\zeta=\frac{\pi^{2}}{2} n_{W}\Delta \lambda^2 {\omega^3\sqrt{\omega^2
-\omega_p^2}  \over cc_s (\omega^2 -\omega_p^2 +c_s^2k_s^2)}~.
                                                  \end{eqnarray}
Thus in the presence of bulk pinning absorption of the EM wave
starts from the threshold frequency $\omega_p$ and above the
threshold grows as $\sqrt {\omega-\omega_p}$. If at the same time
surface pinning is weak enough there is an absorption peak at
$\omega -\omega_p \sim c_sk_s$. The peak can correspond to
frequencies much lower than the threshold frequency for excitation
of bulk spin waves (frequency $\alpha \omega_M$ of the ferromagnetic
resonance) \cite{Braude}. These special features of the absorption
spectrum can be useful for experimental detection of DW waves in
SCFMs.

Our analysis was based on the classical Landau-Lifshitz dynamics of
a spin FM with a single-valued spontaneous magnetization
$\mathbf{M}$. Meanwhile in $p$-wave superconductors
superconductivity coexists with {\em orbital ferromagnetism} when
spontaneous magnetic moment is not well-defined and one should
develop the magnetic dynamics using the concept of magnetization
currents \cite{PW}. But as well as in the case of bulk magnons
considered in Ref. \onlinecite{PW}, one may expect that this would
lead only to redefinition of the 2D magnon parameters without
changing the general picture of their propagation and excitation. We
hope to address this problem elsewhere.

In summary, the low-frequency dynamic of the DW in SCFM has been
studied and the spectrum of DW waves (2D magnons) has been found.
The magnetostatic effects (long-range dipole-dipole interaction)
have a crucial influence on the long-wavelength spectrum, while the
superconducting Meissner currents modify these effects essentially.
They suppress the effect of long range dipole interaction
eliminating the non-analytical features of the 2D-magnon spectrum
revealed for normal FMs. Our analysis has demonstrated that the
response of a SCFM sample to the EM irradiation
(surface impedance) provides information on 2D magnons in DWs. Bulk
and surface pinning of DWs introduces a gap in the 2D-magnon
spectrum and a characteristic peak in the absorption of the incident
EM wave (real part of the surface impedance). In
conclusion, studying  of DW waves promises to be an effective probe
of superconductivity-ferromagnetism coexistence.

This work has been supported by the grant of the Israel Academy of
Sciences and Humanities.


\begin{thebibliography}{99}

\bibitem{ginz} V.~L.~Ginzburg, Zh.~Eksp.~Teor.~Fiz. {\bf 31}, 202 (1956)
 [Sov.~Phys.~JETP {\bf 4}, 153 (1957)].
\bibitem{Felner} I.~Felner {\it et.al.}, Phys. Rev. B {\bf 55}, 3374 (1997).
\bibitem{bernhard} C.~Bernhard {\it et.al.}, Phys. Rev. B {\bf 59}, 14099 (1999).
\bibitem{Eskildsen} M.~R.~Eskildsen {\it et.al.}, Nature (London) {\bf 393}, 242 (1998).
\bibitem{saxena} S.~S.~Saxena {\it et.al.}, Nature (London) {\bf 406}, 587 (2000).
\bibitem{pfleiderer} C.~Pfleiderer {\it et.al.}, Nature (London) {\bf 412}, 58 (2001).
\bibitem{aoki} D.~Aoki {\it et.al.}, Nature (London) {\bf 613}, 413 (2001).
\bibitem{NV} T.K. Ng and C.M. Varma, Phys. Rev. B {\bf 58}, 11624 (1998).
\bibitem{Braude} V.~Braude and E.~B.~Sonin, Phys. Rev. Let. {\bf 93}, 117001
(2003); V.~Braude, Phys. Rev. B {\bf 74}, 054515 (2006).
\bibitem{Sonin} E.~B.~Sonin, Phys. Rev. B {\bf 66}, 100504 (2002).
\bibitem{krey} U. Krey, Intern. J. Magnetism, {\bf 3}, 65 (1972).
\bibitem{Buzdin} M. Faure« and A. I. Buzdin, Phys. Rev. Lett. {\bf 94}, 187202
(2005), {\it ibid.} {\bf 95}, 269702 (2005);  E. B. Sonin, {\it ibid.} {\bf 95},
269701 (2005).
\bibitem{Malozemoff} A.~P.~Malozemoff and J.~C.~Slonczewski, {\it
   Magnetic Domain Walls and in Bubble Materials} (Academic Press, 1979).
\bibitem{LL} L.D. Landau and E.M. Lifshitz, {\it Electrodynamics
of Continuous Media} (Pergamon Press, Oxford, 1984).
\bibitem{Slonczewski} J.~C.~Slonczewski, Int. J. Magn. {\bf 2}, 85 (1972).
\bibitem{PW} V.~Braude and E.~B.~Sonin, Phys. Rev. B {\bf 74}, 064501 (2006).
\end{thebibliography}
  \end{document}